\documentclass[useAMS,usenatbib]{mn2e}

\usepackage{amsmath,psfig,epsfig,graphics}

\def\aap{A\&A}

\def\apj{ApJ}

\def\apjl{ApJ}
\def\mnras{MNRAS}

\def\aj{AJ}

\def\lesssim{\mathrel{\hbox{\rlap{\hbox{\lower4pt\hbox{$\sim$}}}\hbox{$<$}}}}
\def\gesssim{\mathrel{\hbox{\rlap{\hbox{\lower4pt\hbox{$\sim$}}}\hbox{$>$}}}}
\newcommand{\bu}{{\bf u}}
\newcommand{\br}{{\bf r}}
\def\lesssim{\mathrel{\hbox{\rlap{\hbox{\lower4pt\hbox{$\sim$}}}\hbox{$<$}}}}
\def\gesssim{\mathrel{\hbox{\rlap{\hbox{\lower4pt\hbox{$\sim$}}}\hbox{$>$}}}}

\begin{document}

\author[Morandi et al.]
{Andrea Morandi${}^1$\thanks{E-mail: andrea@wise.tau.ac.il}, Marceau
Limousin${}^{2,3}$, Yoel Rephaeli${}^{1,4}$, Keiichi Umetsu${}^{5}$,
\newauthor Rennan Barkana${}^{1}$, Tom Broadhurst${}^{1,6,7}$,
H{\aa}kon Dahle${}^{8}$\\
$^{1}$ Raymond and Beverly Sackler School of Physics and Astronomy,
Tel Aviv University, Tel Aviv, 69978, Israel\\
$^{2}$ Laboratoire d'Astrophysique de Marseille, Universit\'e de
Provence, CNRS, 38 rue Fr\'ed\'eric Joliot-Curie, F-13388 Marseille
Cedex 13, France\\
$^{3}$ Dark Cosmology Centre, Niels Bohr Institute, University of
Copenhagen, Juliane Maries Vej 30, DK-2100 Copenhagen, Denmark\\
$^{4}$ Center for Astrophysics and Space Sciences, University of
California, San Diego, La Jolla, CA 92093, USA\\
$^{5}$ Institute of Astronomy and Astrophysics, Academia Sinica,
P.~O. Box 23-141, Taipei 10617, Taiwan\\
$^{6}$ Department of Theoretical Physics, University of Basque Country
UPV/EHU,Leioa,Spain\\
$^{7}$ IKERBASQUE, Basque Foundation for Science,48011, Bilbao,Spain\\
$^{8}$ Institute of Theoretical Astrophysics, University of Oslo,
P.O. Box 1029, Blindern, N-0315 Oslo, Norway}

\date{}

\title[Triaxiality and non-thermal gas pressure in Abell~1689]
{Triaxiality and non-thermal gas pressure in Abell~1689}
\maketitle

\begin{abstract}

Clusters of galaxies are uniquely important cosmological probes of the
evolution of the large scale structure, whose diagnostic power depends
quite significantly on the ability to reliably determine their masses.
Clusters are typically modeled as spherical systems whose intracluster
gas is in strict hydrostatic equilibrium (i.e., the equilibrium gas
pressure is provided entirely by thermal pressure), with the
gravitational field dominated by dark matter, assumptions that are
only rough approximations. In fact, numerical simulations indicate
that galaxy clusters are typically triaxial, rather than spherical,
and that turbulent gas motions (induced during hierarchical merger
events) provide an appreciable pressure component. Extending our
previous work, we present results of a joint analysis of X-ray, weak
and strong lensing measurements of Abell~1689. The quality of the data
allows us to determine both the triaxial shape of the cluster and the
level of non-thermal pressure that is required if the intracluster gas is
in hydrostatic equilibrium. We find that the dark matter axis ratios
are $1.24\pm 0.13$ and $2.02\pm 0.01$ on the plane of the sky and
along the line of sight, respectively, and that about $20\%$ of the
pressure is non-thermal. Our treatment demonstrates that the dynamical
properties of clusters can be determined in a (mostly) bias-free way, 
enhancing the use of clusters as more precise cosmological probes.
\end{abstract}


\section{Introduction}\label{intro}

Clusters of galaxies are the largest bound systems, formed at
relatively late times. As such, their mass function sensitively
depends on the evolution of the large scale structure (LSS) and on the
basic cosmological parameters. Given this great potential, the use of
clusters as a cosmological probe hinges on our ability to accurately
determine their masses.

Mass determinations based on X-ray observations customarily assume
spherical symmetry and strict hydrostatic equilibrium (HE), i.e.,
that intracluster (IC) gas pressure is provided entirely by thermal
motions, $P_{\rm tot} = P_{\rm th}$.  Under these assumptions X-ray
measurements can be successfully used to constrain the mass profile
\citep{sarazin1988}.

In addition to this X-ray based method, optical observations of giant
arcs and multiple images produced by strong gravitational lensing (SL)
in the central parts of clusters and slight distortions of background
sources in the weak lensing (WL) regime allow us to determine
projected cluster masses without invoking the assumption of
hydrostatic equilibrium \citep{Miralda-Escude1995}, but an a priori
assumption about the spherical symmetry is still often used to deduce
the three-dimensional mass profiles.
 
However, N-body simulations indicate that dark matter (DM) halos
are triaxial with intermediate-major and minor-intermediate axis ratios typically of the order of $\sim 0.8$ \citep[][]{wang2009}, and hydrodynamical numerical simulations 
suggest that even after equilibrium is
established a significant fraction of the pressure support against
gravity comes from subsonic non-thermal gas motions
\citep{lau2009,zhang2010,richard2010,meneghetti2010b}, raising doubts 
on the viability of cluster sphericity and purely thermal gas
pressure.

Moreover, since lensing is sensitive to the integrated mass contrast
along the line of sight, both departures from the spherical assumption
\citep{morandi2010a} and the inclusion of non-thermal pressure support 
\citep{molnar2010} can explain the long-standing discrepancy between cluster masses 
determined from X-ray and gravitational lensing observations, the
latter often being significantly higher than the former.

The main goal of our work is to resolve these discrepancies 
between X-ray and gravitational lensing mass for the the cluster Abell~1689 
\citep[][]{andersson2004,lemze2008,riemer2009,Peng2009,morandi2011a}. 
We tackle this by developing a novel method in order to infer both the
desired triaxial shape and physical properties of clusters and the
non-thermal pressure via a joint X-ray, weak and strong lensing
analysis.

We extend the findings of our previous work on Abell~1689
\citep{morandi2011a}, where we combined X-ray and strong lensing data;
in that work we accounted for the three-dimensional geometry, which
allowed us to resolve the discrepancy between the mass determined from
X-ray and strong gravitational lensing observations in the inner
region ($R\lesssim 400$ kpc), assuming strict HE. In the present paper
we jointly analyze also weak lensing data that map the projected mass
profile out to $\sim 3$ Mpc. With this additional data we are also
able to determine the non-thermal pressure of the IC gas.

Hereafter we assume the flat $\Lambda CDM$ model, with matter density
parameter $\Omega_{m}=0.3$, cosmological constant density parameter
$\Omega_\Lambda=0.7$, and Hubble constant $H_{0}=70 \,{\rm km\;
s^{-1}\; Mpc^{-1}}$. Unless otherwise stated, quoted errors are at the
68.3\% confidence level.

\section{Datasets and analysis}\label{dataan}

A full description of the X-ray and SL analysis can be found in
\cite{morandi2011a}; for details on the WL analysis we refer to
\cite{umetsu2009}. Here we only briefly summarize the most relevant 
or novel aspects of our data reduction and analysis of Abell~1689.

\subsection{X-ray analysis}\label{laoa}

We analyzed two sets of {\it Chandra} observations (ID numbers 6930
and 7289) from the NASA HEASARC archive with a total exposure time of
approximately 150 ks. With respect to \cite{morandi2011a}, we repeated the X-ray data reduction and analysis by implementing the most recent {\it Chandra} calibrations: we used the CIAO software (version 4.3) and the gain file provided within CALDB (version 4.4.3). We measured the gas density profile in a non-parametric way from the surface brightness recovered by a spatial analysis, and we inferred the projected temperature profile by analyzing the spectral data.

The X-ray images were extracted from the two event files in the energy
range $0.5-5.0$ keV, then corrected by the exposure map to remove the
vignetting effects, followed by the masking out of point sources. We
constructed a set of $n=57$ elliptical annuli of minor radius $r_{m}$
around the centroid of the surface brightness with eccentricity
${\epsilon_{b'}(r)}$ fixed to that predicted from the eccentricity
${e_{b'}(r)}$\footnote{$e_{b'}=\sqrt{1-(b'/a')^2}$, $a'$($b'$) being
the major (minor) axis on the plane of the sky.} of the DM halo on the
plane of the sky as determined from the SL data (see Section
\ref{depr} and \citet{morandi2010a}). We then deduced the electron 
density $n_e = n_e(r; {\epsilon_{c'}})$ by deprojecting the
surface brightness profile, obtaining $n=57$ radial measurements in ellipsoidal shells.
Note the dependence of $n_e(r; {\epsilon_{c'}})$ on the eccentricity
${\epsilon_{c'}}$ of the IC gas along the line of sight, still to be
determined \citep[for further details see below and also Appendix~A
of][]{morandi2010a}.

The spectral analysis was performed by extracting the source spectra
from $n^*$ ($n^*=9$) elliptical annuli of minor radius $r^*_{m}$
around the centroid of the surface brightness and with eccentricity as
predicted from that of the the DM halo (as above). For each of the
$n^*$ annuli the spectrum was analyzed by simultaneously fitting
absorbed MEKAL models to the two observations, in the
energy range 0.6-7 keV (0.9-7 keV for the outermost annulus only); we
fixed the redshift to $z=0.183$ and the photoelectric absorption to
the Galactic value. For each of the annuli, we considered three free
parameters in the spectral analysis: the normalization of the thermal
spectrum $K_{\rm i} \propto \int n^2_{\rm e}\, dV$, the
emission-weighted temperature $T^*_{\rm proj,i}$, and the metallicity
$Z_{\rm i}$.

\subsection{Strong lensing analysis}\label{snnen2sl}

For the strong lensing analysis we refer to the findings of
\cite{Limousin2007}, who presented a reconstruction of the mass
distribution of the galaxy cluster Abell~1689 using detected strong lensing
features from deep ACS observations and extensive ground based
spectroscopy. They presented a parametric strong lensing mass
reconstruction using 34 multiply imaged systems, and they inferred two
large-scale dark matter clumps, one associated with the center of the
cluster and the other with a northeastern substructure. We masked out the north-eastern sector of both the 2D projected mass map and the X-ray data, in order to avoid the contribution from this secondary substructure. We masked out also the central 25 kpc, which is affected by the mass distribution of the cD galaxy. From the strong
lensing analysis the major clump of the cluster appears to be
elongated with a major-minor axial ratio on the plane of the sky of
$1.24\pm 0.13$ and a position angle of $0.4\pm 1$ degrees. We rebinned
the two-dimensional projected mass map into elliptical annuli, whose
eccentricity, centroid and position angle are the same as those
inferred from \cite{Limousin2007}. Then we calculated average values
of the elliptical projected mass profile $\Sigma(R)$, $R$ being the
minor radius of the two-dimensional elliptical annuli.  We also
calculated the covariance matrix $\mathbfit{C}_{\rm sl}$ among all the
measurements of $\Sigma(R)$ similarly to the WL analysis (see \S\ref{snnen2sl3} for further details).

\subsection{Weak lensing analysis}\label{snnen2sl3}

For the weak lensing analysis we refer to the findings of
\cite{umetsu2009}, who derived a projected two-dimensional mass map of
Abell~1689 based on an entropy-regularized maximum likelihood combination
of lens magnification with distortion of red background galaxies in
deep Subaru images. Advantages of that approach are: 1) by combining
the distortion and magnification measurements the convergence can be
obtained unambiguously with the correct mass normalization, while the
convergence derived from distortion data alone suffers from the
mass-sheet degeneracy; 2) the method is not restricted to the weak
regime but applies to the whole area outside the tangential critical
curve, where nonlinearity between the surface mass density and the
observables extends out to a radius of a few arcminutes. 
In the two-dimensional projected mass map in WL we masked out the region corresponding to the secondary substructure seen in the SL regime (\S\ref{snnen2sl}), and
then rebinned the data into elliptical annuli, whose eccentricity, centroid, and position angle are the same as that from SL.

In order to calculate average values of the elliptically-symmetric
projected mass profile $\Sigma(R)$, where $R$ is the minor radius of
the two-dimensional elliptical annuli, we compute $\Sigma=\Sigma(R)$
from a weighted radial projection of the two-dimensional surface mass
density ${\bf \Sigma}$. Following \cite{hobson2002}, we can formally
express a linear relation between ${\bf \Sigma}$ and $\Sigma$:
\begin{equation}
{\bf \Sigma} = {\mathcal M}({\bf r})\; \Sigma +{\mathcal N},
\label{eqn:mm1}
\end{equation}
where ${\mathcal N}$ is the noise vector (at the $i$'th pixel) with
covariance matrix $\mathcal{C}$, and ${\mathcal M}={\mathcal M}({ W})$
is the 1D-to-2D mapping matrix. ${\mathcal M}$ is defined as an
$N_{\rm pix} \times N_{\rm ann}$ matrix, with $N_{\rm pix}$ the total
number of pixels, $N_{\rm ann}$ the total number of elliptical annuli,
and elements ${\mathcal M}_i^m=W_{im}$, where $W_{im}$ ($0\le
W_{im}\le 1$) is the fraction of the area of the $i$'th pixel lying
within the $m$'th elliptical bin. An optimal solution for this problem
is given by:
\begin{equation}
\Sigma = ({\mathcal M}^{\rm t}\mathcal{C}^{-1}{\mathcal M})^{-1}
{\mathcal M}^{\rm t}\mathcal{C}^{-1}\;{\bf \Sigma}\ .
\label{eqn:mm3}
\end{equation}

The binned surface mass data $\Sigma$ have covariance matrix
$\mathbfit{C}_{\rm wl}$:
\begin{equation}
\mathbfit{C}_{\rm wl} = ({\mathcal M}^{\rm t}{\mathcal C}^{-1}
{\mathcal M})^{-1} \ .
\label{eqn:mm4}
\end{equation}
We use Monte Carlo integration to calculate $W_{im}$, while we refer $\Sigma$ to the area-weighted minor radius $R_m$ of the $m$'th elliptical bin \citep{umetsu2008}.

Note that unlike \cite{umetsu2008}, who used 2D data in order to retrieve best-fit model parameters, in the present paper we used rebinned surface mass profiles. While we verified that this does not change significantly the desired best-fit model parameter (\S\ref{depr}), fitting rebinned data has the advantage of requiring significant less computational time and it is usually a better approach when it comes to deal with real data, such as WL, possibly affected by systematics. Indeed rebinning involves a smearing of the data and also of their errors (statistical + systematics) on a larger area: provided that the systematics can be regarded as symmetric errors on a large enough area, this is usually a more robust approach than relying on 2D data.

\subsection{Joint X-ray+Lensing analysis}\label{depr}

Here we briefly summarize the major findings of \cite{morandi2010a}
for the joint X-ray+Lensing analysis in order to infer triaxial
physical properties; additional details can be found in
\cite{morandi2007a,morandi2010a}.  We also outline the improvements
implemented in the present work, i.e., the inclusion of WL data and
the measurement of the non-thermal component of the IC gas.

The lensing and the X-ray emission both depend on the properties of
the DM gravitational potential well, the former being a direct probe
of the two-dimensional mass profile and the latter an indirect proxy
of the three-dimensional mass profile through the hydrostatic
equilibrium equation applied to the gas temperature and density. In
order to infer the model parameters of both the IC gas and of the
underlying DM density profile, we perform a joint analysis of SL, WL
and X-ray data. We briefly outline the methodology for inferring
physical properties in triaxial galaxy clusters: (1) We start with a
generalized Navarro, Frenk and White (gNFW) triaxial model of the DM
as described in \cite{jing2002}, which represents the total underlying
mass distribution and depends on a few parameters to be determined,
namely the concentration parameter $c$, the scale radius $r_{\rm s}$,
the inner slope of the DM $\alpha$ and the two axis ratios; (2)
following \cite{lee2003,lee2004}, we recover the gravitational
potential and surface mass profile $\Sigma$ of a dark halo with such a
triaxial density profile; (3) we solve the generalized hydrostatic
equilibrium equation, i.e., including the non-thermal pressure
(Equation~(\ref{aa4}) below), in order to infer the theoretical
three-dimensional temperature profile $T_{\rm gas}$ in a
non-parametric way, given IC gas with a density profile as deprojected
from the X-ray data (see \S2.1) and sitting in the gravitational
potential well previously calculated; finally, (4) the joint
comparison of $T_{\rm gas}$ with the observed temperature and of
$\Sigma$ with the observed surface mass density gives us the
parameters of the triaxial DM density model and the non-thermal
component of the gas, and therefore all the desired physical
properties of the IC gas and the DM triaxial ellipsoids.

In particular, for the X-ray analysis we rely on a generalization of
the hydrostatic equilibrium equation \citep{lau2009,molnar2010}, which
accounts for the non-thermal pressure $P_{\rm nt}$ and reads:
\begin{equation}\label{aa4}
\nabla P_{\rm tot} = -\rho_{\rm gas} \nabla \phi\ ,
\end{equation}
where $\rho_{\rm gas}$ is the gas mass density, $\phi$ is the
gravitational potential, and $P_{\rm tot}= P_{\rm th}+ P_{\rm nt}$.

In this first implementation of our new approach, and given the
limited ability to carry out a more detailed spatial analysis, we
model $P_{\rm nt}$ by assuming that the non-thermal pressure of the
gas is a constant fraction $\xi$ of the total pressure $P_{\rm tot}$,
i.e.,
\begin{equation}
P_{\rm nt}=\xi P_{\rm tot}\ .
\label{pnt12}
\end{equation}
From Equations~(\ref{aa4}) and (\ref{pnt12}) we point out that neglecting
$P_{\rm nt}$ (i.e., setting $P_{\rm tot} = P_{\rm th}$) systematically
biases low the determination of cluster masses, roughly by a factor
of $\xi$. Note that X-ray data probe only the thermal component of the gas $P_{\rm th}=n_e\, {\bf k}  T_{\rm gas}$.

The work of \cite{lee2003} showed that the IC gas and DM halos are
well approximated by a sequence of concentric triaxial distributions
with different eccentricity ratio. We define $e_{b'}$
($\epsilon_{b'}$) and $e_{c'}$ ($\epsilon_{c'}$) as the eccentricity
of DM (IC gas) on the plane of the sky and along the line of sight,
respectively. The iso-potential surfaces of the triaxial dark halo
coincide also with the iso-density (pressure, temperature) surfaces of
the intracluster gas. This is simply a direct consequence of the {\it
X-ray shape theorem} \citep{buote1994}; the hydrostatic equilibrium
equation (\ref{aa4}) yields
\begin{equation}\label{eqn:ecc}
\nabla P \times \nabla\phi = \nabla \rho \times \nabla\phi = 0 .
\end{equation}

Note that $\epsilon_{b'}=\epsilon_{b'}(e_{b'},u,\alpha)$ and
$\epsilon_{c'}=\epsilon_{c'}(e_{c'},u,\alpha)$, with $\bu \equiv
\br/r_{\rm s}$, unlike the constant $e_{b'},e_{c'}$ for the adopted DM
halo profile. In the whole range of $u$, $\epsilon_{b'}/e_{b'}$
($\epsilon_{c'}/e_{c'}$) is less than unity ($\sim 0.7$ at the
center), i.e., the intracluster gas is altogether more spherical than
the underlying DM halo.

We construct the likelihood performing a joint analysis for lensing and X-ray data, in order to constrain the properties of the model parameters ${\bf q}$
\begin{equation}\label{aa334}
{\bf q}=(c,r_{\rm s},\alpha,e_{c'},\xi)\ 
\end{equation}
representing the concentration parameter, scale radius, inner slope of
the DM, eccentricity of the DM along the line of sight and fractional
non-thermal pressure of the gas, respectively. 

Therefore, the system of equation we simultaneously rely on in our joint X-ray+Lensing analysis is:
\begin{subequations}
\begin{equation}\label{5}
T_{\rm gas} = T_{\rm gas}(c,r_{\rm s},\alpha,e_{c'}, \xi)
\end{equation}
\begin{equation}\label{6}
\Sigma = \Sigma(c,r_{\rm s},\alpha,e_{c'})
\end{equation}
\end{subequations}
where the three-dimensional model temperature $T_{\rm gas}$ is recovered by solving equation (\ref{aa4}) and constrained by the observed temperature profile, and the model surface mass profile $\Sigma$ is recovered by projection of the triaxial gNFW DM model and constrained by weak/strong lensing measurements.

The method works by constructing a joint X-ray+Lensing likelihood:
\begin{equation}\label{chi2wwf}
{\mathcal{L}}={\mathcal{L}}_{\rm X}\cdot {\mathcal{L}}_{\rm
lens}={\mathcal{L}}_{\rm X}\cdot {\mathcal{L}}_{\rm SL}\cdot
{\mathcal{L}}_{\rm WL}\ ,
\end{equation}
$\mathcal{L}_{\rm X}$ and ${\mathcal{L}}_{\rm lens}$ being the
likelihoods coming from the X-ray and lensing (WL+SL) data,
respectively. Now, ${\mathcal{L}}_{\rm X}\propto {\exp ( -\chi_{\rm
X}^2/2)}$, where
\begin{equation}\label{chi2wwe}
\chi^2_{\rm X}= \sum_{i=1}^{n^*} {\frac{{ (T_{\rm proj,i}-T^*_{\rm proj,i})}^2 }{\sigma^2_{T^*_{\rm proj,i}}  }}\
\end{equation}
$T^*_{\rm proj,i}$ being the observed projected temperature profile in
the $i$'th ring and $T_{\rm proj,i}({\bf q})$ the convenient
projection \citep[following][]{mazzotta2004} of the theoretical
three-dimensional temperature $T_{\rm gas}({\bf q})$; the latter is
the result of solving the hydrostatic equilibrium equation, with the
gas density $n_e(r; {\epsilon_{c'}})$ inferred from the X-ray surface
brightness, and assuming a gNFW parametrization for the DM $\rho_{\rm
DM}=\rho_{\rm DM}({\bf r}; c,r_{\rm
s},\alpha,e_{c'})$. The lensing contribution is
\begin{equation}\label{aa2w2q}
\!{\mathcal{L}}_{\rm lens}=\frac{\exp \left\{-\frac{1}{2}
{[ \Sigma-\Sigma^*]}^{\rm t}\mathbfit{C}^{-1} [
\Sigma-\Sigma^*]\right\}} {(2\pi)^{m^*/2}|\mathbfit{C}|^{1/2}}\ ,
\end{equation}
where $\mathbfit{C}$ is the covariance matrix of the projected mass
profile from strong ($\mathbfit{C}_{\rm sl}$) and weak
($\mathbfit{C}_{\rm wl}$) lensing data, $|\mathbfit{C}|$ indicates the
determinant of $\mathbfit{C}$,
$\Sigma^*=(\Sigma_1^*,\Sigma_2^*,...,\Sigma_{m^*}^*)$ are the observed
measurements of the projected mass profile in the $m^*$ elliptical
annuli, and $\Sigma(c,r_{\rm s},\alpha,e_{c'})$ is the theoretical
projected mass profile within our triaxial DM model.

For the covariance matrix of the strong lensing measurements, we also 
account for systematic errors by replacing the observed
covariance matrix with the following expression:
\begin{equation}\label{aartb}
\mathbfit{C}_{\rm sl}\longrightarrow{\mathbfit{C}_{\rm sl}}\; + \sigma^2_{\rm sys}\; {\mathcal{I}}
\end{equation}
where ${\mathcal{I}}$ is the identity matrix. Since we are using the
surface mass density as output from the parametric SL analysis of
\cite{Limousin2007}, we introduce the parameter
$\sigma_{\rm sys}$ to allow some extra freedom due to possible
systematic bias resulting from their analysis. In order to calculate $\sigma_{\rm sys}$ we assumed that systematic errors can be described as Gaussian errors via a diagonal matrix $\sigma^2_{\rm sys}\; {\mathcal{I}}$ with the same value in each of the diagonal elements. We checked that this simplified assumption does not significantly affect the average value of the physical parameters, while it slightly increases (12\%-20\%) their errors with respect to the case where we neglect $\sigma_{\rm sys}$.

We evaluated the probability distribution function of model parameters
${\mathcal{L}}={\mathcal{L}}({\bf q}, \sigma_{\rm sys})$ via the
Markov Chain Monte Carlo (MCMC) algorithm. Thus we can determine our
best-fit model parameters and their errors, and derive various
properties of the cluster.

\section{Results and Discussion}\label{dataan2}

In the previous section we showed how we can determine the physical
parameters of the cluster by fitting the available data based on the
hydrostatic equilibrium equation and on a DM model that is based on
robust results of hydrodynamical cluster simulations. In this section
we present our results and discuss their main implications. We
particularly focus on the implications of our analysis for the
viability of the CDM scenario, the discrepancy between X-ray and
lensing masses in Abell~1689, and the presence of non-thermal pressure.

\subsection{Best-fit parameters}

\begin{figure*}
\begin{center}
\psfig{figure=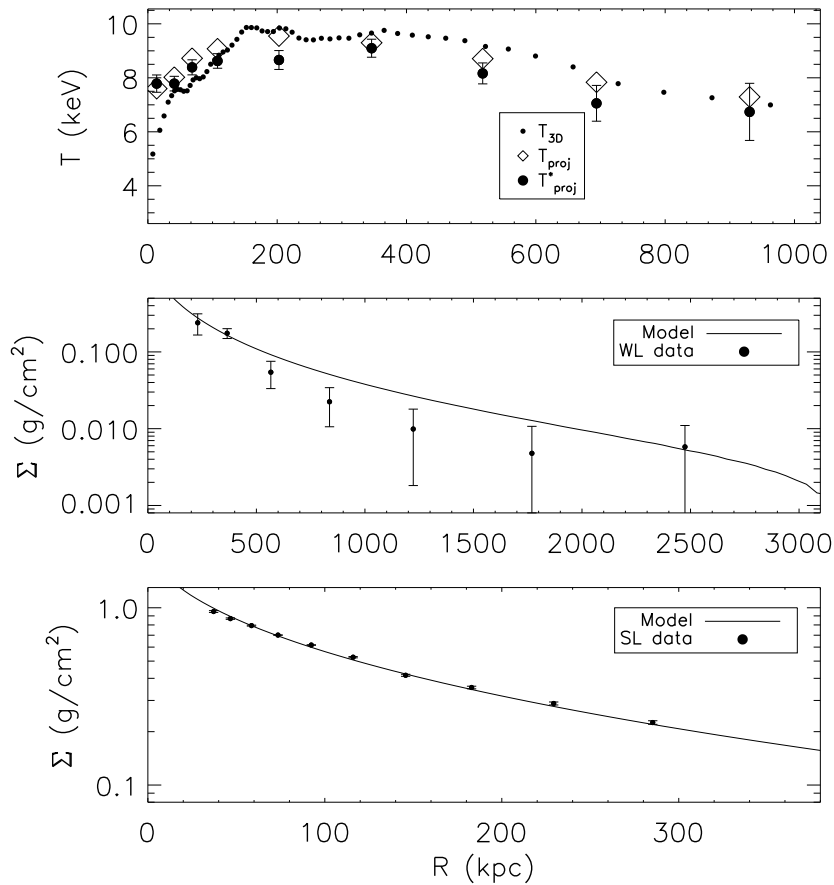,width=0.7\textwidth}
\caption[]{Joint X-ray and lensing analysis for Abell~1689. In the upper 
panel we display the two quantities that enter in the X-ray analysis
(Equation (\ref{chi2wwe})): the observed projected
temperature $T^*_{\rm proj,i}$ (big points with error bars) and the
theoretical projected temperature $T_{\rm proj,i}({\bf q})$
(diamonds). We also show the theoretical three-dimensional temperature
$T_j({\bf q})$ (points), which generates $T_{\rm proj,i}({\bf q})$
through convenient projection techniques. In the middle and lower
panel we display the two quantities which enter in the WL and SL
analysis, respectively (Equation (\ref{aa2w2q})): the observed surface
mass profile $\Sigma_i^*$ (points with error bars) and the theoretical
one $\Sigma({\bf q})$ (solid line). The distance of the points for
both $T_j({\bf q})$ and $\Sigma_i^*$ is representative of the true
spatial resolution of X-ray and lensing data, respectively.}
\label{entps332333}
\end{center}
\end{figure*}

In Figure~\ref{entps332333} we present the results of our joint
analysis for Abell~1689. Both the X-ray and lensing data are reasonably
fitted by our model, with a total $\chi^2_{\rm tot}=\chi^2_{\rm
X}+\chi^2_{\rm sl}+\chi^2_{\rm wl}=33.8$ (with 20 degrees of freedom).
The separate contributions are $\chi^2=12.4,6.3,15.1$ for X-ray, SL
and WL data, respectively, where $\chi^2_{\rm lens}\propto
-2\log({\mathcal{L}}_{\rm lens})$.

Note that in Figure~\ref{entps332333} the WL model predictions are
slightly in disagreement with the data, overpredicting the surface
mass profile in the range 500-2000 kpc at around the 1-$\sigma$
significance level, while the X-ray and SL data are rather well
fitted. We checked that the inclusion of WL data does not significantly
change the values of the desired physical parameters. Also, we
verified that the joint fit is preferentially more strongly
constrained by the X-ray and SL data with the present datasets, WL
measurements being of poorer quality. Excluding WL data slightly 
increases (5\%-15\%) the errors on the final parameters with respect to 
the case where we include them, with the exception of $P_{\rm nt}$, where the error would increase of $\sim$ 60\%. With this respect, while for the present datasets WL data are less relevant in constraining the model parameters with respect to X-ray and SL data, whose quality is  exceptionally good, we observe that this might not apply for other galaxy clusters and other datasets, where the inclusion of WL data could be important in removing the degeneracy among the same parameters.

To further probe the
disagreement between WL and SL data, we jointly fitted them (without the X-ray data)
with a spherical NFW ($\alpha=1$) model first (following
\cite{umetsu2008}), and then with the more general triaxial gNFW. We
found a marginal disagreement between SL and WL data for both
models. A similar slight disagreement can also be seen in Figure~13 of
\cite{umetsu2008}.
 
In order to make a comparison with \cite{morandi2011a}, we also compared the $\chi^2$ by neglecting WL data and without $P_{\rm nt}$: we find $\chi^2=25.3$ (14 degrees of freedom), while in \cite{morandi2011a} we found that the $\chi^2\sim 7.4$ (11 degrees of freedom). Note that the difference in the value of the $\chi^2$ reflects a different treatment of the SL data in the present paper (\S\ref{snnen2sl} and \S\ref{snnen2sl3}) in comparison to our previous work. Among these difference, the major improvement is based on the determination of covariance ${\mathcal C}_{\rm sl}$ among the pixels of the 2D map distribution, in order to infer the covariance $\mathbfit{C}_{\rm sl}$ among the rebinned measurements (Equation \ref{eqn:mm4}), and on an improved rebinning schema (Equation \ref{eqn:mm3}). We also point out that this difference in the value of the $\chi^2$ translates to slightly different errors in the parameter values, with negligible impact on their expectation values. For example, we find that the axis ratio along the line of sight $\eta_{\rm{DM},c'}=2.39\pm 0.02$ (by neglecting WL data and without $P_{\rm nt}$), whereas the value deduced in \cite{morandi2011a} was $\eta_{\rm{DM},c'}=2.37\pm 0.11$.

In Table~\ref{tabdon} we present the best-fit model parameters for our
analysis of Abell~1689. Errors in the individual parameters $({\bf q},
\sigma_{\rm sys})$ have been evaluated by considering the average value
and the mean absolute deviation of the marginal probability
distribution of each parameter. Note that we list the axis ratio along
the line of sight $\eta_{\rm{DM},c'}$ rather than the
eccentricity. Our work shows that Abell~1689 is a triaxial galaxy cluster
with DM halo axis ratios $\eta_{\rm{DM},b'}=1.24\pm 0.13$ and
$\eta_{\rm{DM},c'}=2.02\pm 0.01$, where $\eta_{\rm{DM},b'}$ is the
axis ratio on the plane of the sky inferred from SL measurements, and
$\eta_{\rm{DM},c'}$ is the axis ratio along the line of sight inferred
through our joint analysis. Note that these elongations are
statistically significant, i.e., our results disprove the spherical
geometry assumption. The
axis ratio of the gas is $\eta_{\rm{gas},b'}\sim 1.1-1.06$ (on the
plane of the sky) and $\eta_{\rm{gas},c'}\sim 1.5-1.3$ (along the
line of sight), moving from the center toward the X-ray boundary.

In Table~\ref{tabdon} we also report the value of $M_{200}$:
\begin{equation}\label{aartb55}
M_{200} = \frac{800\pi}{3}(a_r/c_r)(b_r/c_r)R_{200}^3 \, \rho_c \ ,
\end{equation}
where $\rho_c$ is the critical density of the Universe at the redshift $z$ of the cluster, $c = {R_{200}}/{r_{\rm s}}$, and $a_r/c_r$ and $b_r/c_r$ are the minor-major and intermediate-major axis ratios of the DM halo, respectively.


\begin{table*}
\begin{center}
\caption{Best-fit model parameters of A1689. The columns $1\!\!-\!6$ refer to the best fit parameters $c,r_{\rm s},\alpha, \eta_{DM,c'}$, $\xi$ and $\sigma_{\rm sys}$, while the last column refers to $M_{200}$.}
\begin{tabular}{c@{\hspace{.7em}} c@{\hspace{.7em}} c@{\hspace{.7em}} c@{\hspace{.7em}} c@{\hspace{.7em}} c@{\hspace{.7em}} c@{\hspace{.7em}}}
\hline \\
 $c$ & $r_{\rm s}$  & $\alpha$ & $\eta_{DM,c'}$ & $\xi$ & $\sigma_{\rm sys}$ & $M_{200}$\\
     &     (kpc)    &           &       &       & ${(\rm g\; cm^{-2})}$ & ($10^{15}M_{\odot}$)\\
\hline \\
  $5.71\pm 0.47$ & $348\pm 28$  &  $0.95\pm 0.05$  &  $2.02\pm 0.01$ & $0.22\pm 0.01$ & $0.008 \pm 0.001$ & $2.59\pm0.09$\\
\hline \\\\
\end{tabular}
\label{tabdon}
\end{center}
\end{table*}

The second main result from our work is the need for a appreciable
non-thermal pressure support, formally at a level of $\sim$20\%.

\begin{figure}
\begin{center}
\psfig{figure=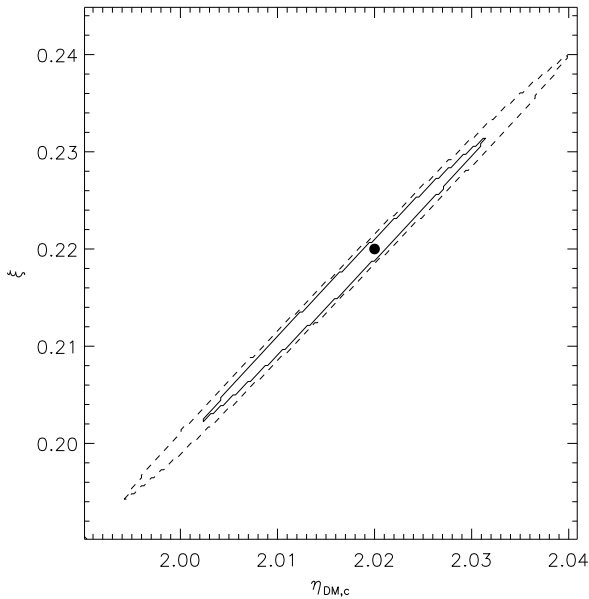,width=0.48\textwidth}
\caption[]{Joint probability distribution of  $\eta_{\rm{DM},c'}$ and 
$\xi$. The solid(dashed) line represents the 1(2)-$\sigma$ error
region, while the big dot represents the best-fit values.}
\label{entps3xkn}
\end{center}
\end{figure}

In Figure~\ref{entps3xkn} we present the joint probability
distribution of $\eta_{\rm{DM},c'}$ and $\xi$. Note the positive correlation between $\xi$ and $\eta_{\rm{DM},c'}$, i.e. the X-ray/Lensing mass discrepancy in clusters with prominent strong lensing features can be explained via a combination of both triaxiality and non-thermal support. This expected in light of the following considerations: (1) the observed temperature profile and projected mass profile are both sensitive to triaxiality, specifically $T(R,\eta_{\rm
DM}) \sim \eta_{\rm{gas},c'} \, T(R,\eta_{\rm DM}=1)$ and $\Sigma(R)
\propto \eta_{\rm{DM},c'}$ \citep{morandi2010a}, so that the dependency 
of $\Sigma(R)$ on triaxiality is stronger than that of $T(R)$
(remembering that generally $\eta_{\rm{DM},c'} > \eta_{\rm{gas},c'}$);
(2) the non-thermal pressure support affects only the X-ray data via
Equation~\ref{aa4}, i.e., neglecting $P_{\rm nt}$ systematically
lowers the determination of cluster masses based on X-ray data roughly by the
factor $\xi$.

\subsection{Implications for the viability of the CDM scenario 
and the X-ray/Lensing mass discrepancy}\label{conclusion33a}

A precise determination of cluster masses is crucial for the use of
clusters as cosmological probes. However, there are discrepancies
between cluster masses determined based on gravitational lensing and
on X-ray observations, the former being significantly higher than the
latter in many clusters with prominent lensing features. Indeed,
\cite{Oguri2009} showed that SL clusters with the largest Einstein
radii constitute a highly biased population with major axes
preferentially aligned with the line of sight thus increasing the
magnitude of lensing. Given that lensing depends on the
integrated mass along the line of sight, either fortuitous alignments
with mass concentrations that are not physically related to the galaxy
cluster or departures of the DM halo from spherical symmetry can bias
upwards the three-dimensional mass with respect to the X-ray mass
\citep{Gavazzi2005}; on the other hand, X-ray-only masses hinge on
the accuracy of the assumption of strict hydrostatic equilibrium: the
presence of bulk motions in the gas can bias low the three-dimensional
mass profile between 5 and 20\% \citep{meneghetti2010b}.

\cite{Oguri2005} concluded that weak lensing measurements in Abell~1689 
are indeed compatible with the CDM-based triaxial halo model if Abell~1689
represents a rare population ($\sim$6\% by number) of cluster-scale
halos, and \cite{morandi2011a} demonstrated that triaxiality allows us
to remove the mass discrepancy between the strong lensing and X-ray
estimates in Abell~1689. \cite{molnar2010} suggested an alternative
explanation for the mass discrepancy, due to violations of strict
hydrostatic equilibrium in Abell~1689: they found that a contribution of
about 40\% from non-thermal pressure within the core region of Abell~1689
can explain the mass discrepancy, provided that the spherical geometry
assumption holds.

Moreover, recent work investigating the mass distributions of
individual galaxy clusters (Abell~1689 and others) based on gravitational
lensing and employing standard spherical modeling have found a
potential inconsistency compared to the predictions of the CDM
scenario relating halo mass to the concentration parameter $c$. In
particular, relatively high values of $c$ ($\sim 8$-$14$) have been
derived from lensing analysis of Abell~1689
\citep{broadhurst2005,Limousin2007}. These values are outside the
range predicted from simulations of the standard CDM model
\citep[$c\sim 4$;][]{neto2007}. 

In the present work we presented a physical model for the cluster
Abell~1689 with a triaxial mass distribution including support from
non-thermal pressure. This model is consistent with X-ray and SL 
observations as well as with the predictions of the CDM
model. Doing so we have removed the apparent discrepancy between X-ray
and lensing mass estimates, and potential 
inconsistencies discussed in the literature between the predictions of
the CDM scenario and the measurements in clusters with prominent
strong lensing features. One of the main results of our work is the
measurement of a central slope of the DM profile $0.95\pm 0.05$ in
agreement with the theoretical expectation of the CDM scenario
\citep{navarro1997,Merritt2006}, and a value of the concentration
parameter $5.71\pm 0.47$, in agreement with the theoretical
expectation from N-body simulations of \cite{neto2007}, where
$c\sim 4$ at the redshift and for the virial mass of Abell~1689, with an
intrinsic scatter of $\sim 20$\%. In this way we also managed to
reproduce the large Einstein radius of Abell~1689 via our triaxial
framework \citep[][]{morandi2011a}.

Even with triaxiality, though, \cite{Oguri2005} found that Abell~1689 is
relatively rare in its lensing efficiency, as noted
above. Furthermore, \cite{Broadhurst2008} showed that the observed
Einstein radius $\theta_{\rm E}$ of four well-studied massive clusters
(including Abell~1689) lies well beyond the predicted distribution of
Einstein radii in the standard CDM model, typically by a factor of 2,
even after accounting for the actual projected mass distributions in
N-body simulations, and for lensing bias. Using a larger sample of 12
X-ray selected clusters, \cite{Zitrin2011} found a smaller discrepancy
with CDM predictions, typically by a factor of 1.4. 

In a similar fashion, by comparing numerical and observational
cluster samples, \cite{meneghetti2011a} show that some real clusters have
too large lensing cross sections and Einstein rings compared
to expectations in a $\Lambda CDM$ cosmological model. \cite{horesh2011} partially confirmed these results, by comparing the lensed arc statistics measured from the Millennium simulation to those of a sample of observed clusters: they found an excellent agreement between the observed and simulated  number of arcs in the redshift range $0.3<z<0.5$, while at lower redshift some conflict still remains, with real clusters being $\sim 3$ times more efficient arc producers than their simulated counterparts. This departure might be due to selection biases in the observed subsample at this redshift or to physical effects that arise at low redshift and enhance the lensing efficiency, and not included the simulations, as the effect of baryons on the DM profile in clusters \citep{BL2010}.

We also compare the major-minor principal axis ratio
$\eta_{\rm{DM},c'}=2.02\pm 0.01$ with that inferred in our previous
work \citep[$\eta_{\rm{DM},c'}=2.37\pm 0.11$; ][]{morandi2011a}. The
estimated value of the elongation along the line of sight in the
present work is somewhat smaller, due to the assumption of strict
hydrostatic equilibrium and the lack of the large-scale WL data in our
previous work. This lends further support to our emphasis on 
the role of the effects of both geometry and non-thermal pressure support on the physical parameters and open a new window in recovering the intrinsic shapes and desired physical parameters of galaxy clusters in a bias-free way. The estimated value
of $\eta_{\rm{DM},c'}$ is also consistent with the results from
numerical simulations \citep{Shaw2006}.

\subsection{Non-thermal gas pressure}\label{conclusion33b}

Simulations of galaxy clusters predict that turbulent motions should
occur in the IC gas while the matter continues to accrete along
filaments. This energy should then cascade from large to small scales
and can eventually dissipate into the gas. The attainment of HE does
not by itself rule out an appreciable level of turbulence
\citep[$\sim$5-15\%, ][]{lau2009,zhang2010}. Measurements 
of the non-thermal energy in the IC gas are important in order to
estimate the amount of energy injected into clusters from mergers,
accretion of material or feedback from active galactic nuclei (AGNs);
in the latter case, this probes energy transport from the central
nucleus into its surroundings and into the IC gas. \cite{bruggen2005}
found that in clusters with AGN feedback gas motion induced by the
inflation of bubbles and their buoyant rise leads to velocities of
about 500$-$1000 ${\rm km\; s^{-1}}$, which is a significant fraction
of the local sound speed. Strong turbulent motions would provide
significant non-thermal pressure support; this would bias low the
determination of cluster mass profiles measured under the assumption
that $P_{\rm tot} = P_{\rm th}$ in Equation (\ref{aa4}). The
measurement of turbulence is thus important for mass determination as
well as understanding AGN feedback. Moreover, \cite{molnar2010} argued
that there is no need for feedback from a central AGN for the strict
hydrostatic equilibrium to break down in the central regions of
clusters: subsonic random gas motions, a direct consequence of
hierarchical structure formation, allow us to explain violations of
the strict hydrostatic equilibrium assumption in the central regions;
in the outer regions violations could be due to more recent and still
ongoing slow accretion, since those regions have not reached
equilibrium due to the large sound crossing time.

Observationally, \cite{richard2010} measured an X-ray/Lensing mass discrepancy of 
$<M_{\rm sl}/M_{\rm x}>=1.3$ at 3-$\sigma$ significance level in a sample of 20 
strong lensing clusters. They interpret this as evidence that the assumption 
of strict hydrostatic equilibrium required by the X-ray mass estimates is not wholly 
reliable and the merging activity can add non-thermal pressure support 
to the IC gas through bulk motions. 
\cite{sanders2010a} placed a direct limit on turbulence based
on the non-thermal velocity broadening measured from the emission
lines originating in the central 30 kpc of the galaxy cluster
Abell~1835. They found that the ratio of turbulent to thermal energy
density in the core is less than 13\%.

\cite{molnar2010} analyzed a sample of massive clusters of 
galaxies drawn from high-resolution cosmological simulations and found
a significant contribution (20\%-45\%) from non-thermal pressure. They
also tested the validity of strict hydrostatic equilibrium in Abell~1689
using gravitational lensing and X-ray observations under the
assumption of spherical geometry in order to explain the X-ray/Lensing
mass discrepancy: they found a contribution of about 40\% from
non-thermal pressure within the core region of Abell~1689, suggesting an
alternate explanation for the mass discrepancy, as long as the
spherical assumption holds.

Our results point to a scenario where the non-thermal component is
about $20\%$ of the total energy budget of the IC gas. 
This level is quite lower than that found from
\cite{molnar2010} under the assumption of spherical geometry,
suggesting that accounting for the proper triaxial geometry is quite
important in evaluating $P_{\rm nt}$ accurately. 

Another relevant
consideration here is that in our model we constrained the non-thermal
pressure to be a constant factor of the local thermal pressure
throughout the cluster. The results of \cite{mahdavi2008} show that there is a radial trend of the X-ray/WL mass ratio, that is interpreted as
caused by non-thermality increasing toward the outer regions, though their findings hinge on the assumed spherical geometry, so they did not disentangle
the effect of triaxiality from non-thermal pressure support. In this perspective, we
used a linear relation of $P_{\rm nt}$, by fixing the slope to that found by \cite{mahdavi2008} and leaving
the normalization as a free parameter: the results are not affected appreciably from this, the reason is that the fit is more dominated by the innermost part ($<300$ kpc) of the Xray and SL data. Moreover, in order to gauge the likely model dependence of the inferred
non-thermal pressure, we performed a joint analysis by excluding the
temperature constraints (Equation \ref{chi2wwe}) and assuming a
spherical NFW model for the DM. Doing so also allows us to make a
direct comparison with \cite{lemze2008}, who used a similar approach.
We found that the ratio between the (de)-projected temperature profile
measured from the X-ray spectrum and that obtained by assuming strict
hydrostatic equilibrium (Equation \ref{aa4} with $P_{\rm tot}= P_{\rm
th}$) is $\sim0.5-0.6$, which is lower than the determination in our
triaxial model ($P_{\rm th}/P_{\rm tot}\sim 0.8$). Note that
\cite{lemze2008} concluded that a factor of 0.7 explains most of the
temperature discrepancy in Abell~1689, whereas we find a somewhat lower
value; this stems mostly from the more recent Chandra
calibrations adopted in the present work, 
which led to a lower spectral temperature profile (by
$\sim 20\%$). While this test heavily hinges on the model of spherical
geometry, it suggests that the small formal error on $\xi$
in our triaxial joint analysis merits some caution. 

\section{Summary and conclusions}\label{conclusion33}

In this paper we have employed a physical cluster model for Abell~1689 with
a triaxial mass distribution including support from non-thermal
pressure, and proved that it is consistent with the X-ray and SL observations and the predictions of CDM models.

We demonstrated that accounting for the three-dimensional geometry and
the non-thermal component of the gas allows us to resolve the
long-standing discrepancy between the X-ray and strong lensing mass of
Abell~1689 in the literature, as well as to measure a central slope of the
DM and a concentration parameter in agreement with the
theoretical expectations of the CDM scenario.

We also measured the contribution of the non-thermal component of the
gas ($\sim 20\%$ of the total energy budget of the IC gas). This has
important consequences for estimating the amount of energy injected
into clusters from mergers, accretion of material or feedback from
AGN.

The increasing precision of observations now makes it possible to test
the assumptions of spherical symmetry and hydrostatic equilibrium.
Since important current cosmological tests are based on the knowledge
of the masses, shapes, and profiles of galaxy clusters, it is
important to better characterize their physical properties by allowing
for realistic triaxial structures as well as non-thermal
pressure support. The application of our method to 
a larger sample of clusters 
will allow to infer the desired physical parameters of galaxy clusters in a 
bias-free way, with important implications on the use of galaxy clusters as precise cosmological probes.

\section*{acknowledgements}
A.M. and R.B. acknowledge support by Israel Science Foundation grant
823/09. M.L. acknowledges the Centre National de la Recherche
Scientifique (CNRS) for its support. The Dark Cosmology Centre is funded by the Danish National Research Foundation.


\newcommand{\noopsort}[1]{}

\end{document}